\documentclass[12pt,a4paper,final]{iopart}

\usepackage{natbib}
\usepackage{amsfonts}
\usepackage{bm}
\usepackage{iopams}  
\usepackage{graphicx}
\usepackage[breaklinks=true,colorlinks=true,linkcolor=blue,urlcolor=blue,citecolor=blue]{hyperref}

\newcommand{\bb}{\mathbf}

\newcommand{\beq}{\begin{equation}}
\newcommand{\eeq}{\end{equation}}
\newcommand{\bed}{\begin{displaymath}}
\newcommand{\eed}{\end{displaymath}}
\def\bea{\begin{eqnarray}}
\def\eea{\end{eqnarray}}
\newcommand{\veps}{\varepsilon}

\begin{document}

\title{Simulating (p)reheating after inflation via the DCE?}

\author{Wade Naylor}
\address{Interdisciplinary Research Building, International College and Department of Physics, Osaka University, Toyonaka, Osaka 560-0043, Japan}
\ead{naylor@phys.sci.osaka-u.ac.jp}

\begin{abstract}
We note some close parallels between preheating/perturbative reheating, (p)reheating, models in post-inflationary cosmology and the dynamical Casimir effect (DCE) in quantum optics. For the plasma-mirror model we show how the effective plasma mass (arising from conduction electrons) behaves like an oscillating inflaton field, while created photons behave like a scalar field coupled quadratically to the inflaton. Furthermore, the effect of spacetime expansion can also be incorporated by varying the dielectric function. We propose an experiment that could mimic (p)reheating for both narrow and broad parametric resonance, by employing technology already being used in attempts to detect DCE photons via plasma-mirrors.
\end{abstract}

\vspace{2pc}
\\
\\
\submitto{\JPCM}
\maketitle

\section{A brief introduction to (p)reheating}
\label{intro}

\par The inflationary paradigm is the leading theory which solves the horizon and flatness problems in big-bang cosmology and also predicts a {\it natural} mechanism for quantum to classical freezing of the fluctuations leading to density perturbations, e.g., see \cite{Mukhanov:2005}.\footnote{The full details of this mechanism are still being debated; however, their are analogies with quantum optics via squeezing, e.g., see \cite{Polarski:1995jg}.}  However, the theory of preheating/perturbative reheating (denoted (p)reheating from now on): how matter is created after inflation still remains incomplete, although much progress has been made since the works of \cite{Dolgov:1989,Traschen:1990, Shtanov:1994, Kofman:1997}, for a recent review, see e.g., \cite{ReheatRev}.

\par  In cosmology we are fortunate to be able to work with the Friedman-Lema\^itre-Robertson-Walker (FLRW) metric 
defined by
\begin{equation}
ds^{2}=-dt^{2}+a^{2}(t)d\vec{x}^{2}~,
\label{FLRW}
\end{equation}
where $a(t)$ is known as the scale factor, which (depending on the form) stretches or shrinks a local patch of Minkowski space. Later we will see how we can associate the scale factor with the dielectric constant in an optical system: $a(t) \equiv\veps(t)$. Hence there is the possibility of observing interesting time-dependent phenomena in dielectrics (quantum optics) that literally open windows to understanding (p)reheating in cosmology. Note that in cosmology it is standard to work in units where $\hbar=c=1$. 


\par The theory of perturbative reheating (and also preheating) after inflation derives from following observations. Consider the following toy Lagrangian:
\beq
{\cal L}(\phi,\chi)=\frac{1}{2}\partial_{\mu}\phi \partial^{\mu}\phi +\frac{1}{2}\partial_{\mu}\chi \partial^{\mu}\chi 
-V(\phi) -\frac{1}{2}m_{\chi}^{2}\chi^{2} -\frac{1}{2}g^{2}\phi^{2}\chi^{2}~,
\label{toyL}
\eeq
where for simplicity we ignore fermionic couplings implying  only the creation of scalar matter and we assume $V(\phi)=1/2 m^2 \phi^2$, which is symmetric about the origin and therefore only has quadratic couplings.\footnote{See e.g. \cite{Kofman:1997} for a discussion of how symmetry breaking potentials lead to cubic interaction terms, e.g. $\phi \chi^2$.} After inflation ends, where the Hubble constant is assumed much smaller than the inflaton mass: $H \ll m$ the EOM becomes (for a homogeneous field)
\begin{equation}
\ddot{\phi}+3H\dot{\phi}+\Gamma\dot{\phi}+m^{2}\phi=0~,
\end{equation}
where the third term arises from the interaction term, $\frac{1}{2}g^{2}\phi^{2}\chi^{2}$, in Eq. (\ref{toyL}). This term is derived perturbatively from the imaginary part of the effective action for one-particle scattering processes, see \cite{Kofman:1997} . Because of these terms, the amplitude of the sinusoidal oscillations of the inflation $\phi$ decrease as $e^{-\frac{1}{2}\left(3H+\Gamma\right) t}$. The above effective equation only applies when the amplitude, $\Phi$, is relatively small, and is called the elementary theory of perturbative reheating. 

\par Whether or not the universe thermalises depends on the estimated decay rate (\cite{Kofman:1997}):
\begin{equation}
\Gamma(\phi\phi\longrightarrow \chi\chi)={g^4\Phi^2\over 8\pi m}
\label{pertG}
\end{equation}
being larger than the Hubble radius  ($\Gamma > H$). However $H\propto t^{-1}$ (for a matter-dominated phase with $a\propto t^{2/3}$) and the decay rate above is proportional $\Phi^2 \propto t^{-2}$ and therefore the inflaton field never decays fully and perturbative reheating never completes, unless we resort to symmetry breaking potentials, e.g., see \cite{Kofman:1997}. The decay rate above only depend on coupling parameters and mass terms; however, the idea of preheating is that we already have a large condensate of previously produced $\chi$ particles and hence the decay rate will depends on the previous particle number (see Sec. \ref{create}). When the oscillations of the inflaton field are large we can generate an explosive amount of $\chi$ scalar fields through many-particle processes that arise from parametric enhancement. This idea is known as preheating.\footnote{The ``p" can be thought of as for ``pre", as for before perturbative reheating, or for ``parametric" (explosive) particle creation; while it can also be though of as for that in ``perturbative" reheating.}

\par Before we observe this fact (see Sec. \ref{create}) we now treat $\chi$ as a quantum field using the Heisenberg representation, $\hat{\chi}(\vec{x},t)$, and expand it in terms of creation and annihilation operators:
\begin{equation}
\hat{\chi}(\vec{x},t)=\frac{1}{(2\pi)^{{3/2}}} \int d^{3}k \left( \hat{a}_{\vec{k}}(t) e^{i\vec{k}\cdot \vec{x}}
                        \chi_k(t)+h.c.\right). 
\end{equation}
The EOM for $\chi_k$ for each $\vec{k}$ is then 
\begin{equation}
\ddot{\chi}_{k}+3\frac{\dot{a}}{a}\dot{\chi}_k+\left(\frac{\vec{k}^{2}}{a^{2}}+m_{\chi}^{2}+g^{2}\phi^{2} 
        \right)\chi_{k}=0
\end{equation}
which for later purposes, by redefining as $X_k(t)=a^{3/2}(t) \chi_k(t)$, the EOM can be written as
\beq
\ddot X_k+ \omega_k^2(t) X_k =0 
\label{PMathieu}
\eeq
where the effective frequency is
\beq
\omega_k^2= {{k}^{2}\over a^{2}(t)}+g^2\Phi^2 \sin^2 (mt)+m_{\chi}^{2}-\frac{3}{4}\left(\frac{\dot{a}}{a}\right)^{2}-\frac{3}{2}\frac{\ddot{a}}{a}-\xi R
\label{freqP}
\eeq
and for later discussions we have also included a non-minimal coupling term ($\frac 1 2 \xi\phi^2 R$ added to Eq. (\ref{toyL})), where for an FLRW metric
\beq
R = 6[(\dot a/a)^2 + \ddot a/a]
\label{curv}
\eeq 
and $m_\chi$ is the mass of the $\chi$ field. The last two terms are normally neglected in preheating. However, for any cosmological background the latter terms (with $\dot a, \ddot a$) can also contribute to particle creation in general and in models where a shift symmetry protects fermionic (p)reheating, then this may be the only way to thermalise the universe into fermionic matter, e.g.  see \cite{ford1987gravitational}.

\par We will discuss particle creation from Eq. (\ref{PMathieu}), which is a Mathieu-like equation, e.g., see \cite{MathieuBook}, in Sec. \ref{create}. Before that, we will compare this equation with an equation very similar in form that appears in the dynamical Casimir effect (DCE): via varying in time the properties of dielectric media, \cite{DodonovPRA47}. For recent reviews of the DCE, see \cite{Dodonov:2010zza} and \cite{Dalvit:2011yua}.

\section{The DCE in quantum optics}

\par Maxwell's equations including a plasma-mirror term can  be derived from the following Lagrangian (\cite{Yamamoto:2011pra}):
\beq
{\cal L} = \frac 1 2   \veps(t)  \left({\partial\over \partial t}\Psi\right)^2 -\frac 1 2{1\over\mu(t)} (\nabla \Psi )^2 - \frac 1 2 m_p^2(t)\Psi^2
\label{Lag}
\eeq
($\veps_0=\mu_0=1$), where we have denoted the TE field by $\Psi$. In the above we assume that the electric permittivity and magnetic permeability are time dependent, but piecewise constant in space: $\nabla \mu=\nabla \veps=0$. The TM field $\Phi$ (not to be confused with the inflaton amplitude) is obtained by swapping $\veps \leftrightarrow \mu$ in the above Lagrangian, e.g., see \cite{Naylor:2013} (in what follows we shall set $\mu=1$). The plasma mass term, $m_p^2(t)$, represents a varying (real) conductivity in the plasma-mirror model as can be seen heuristically by considering the Drude model of a dielectric with a frequency dependent part:
\beq
\veps(\omega)=\veps_i(t)\left[1-\left( {\omega_p^2\over \omega^2} \right)\right]~,\qquad\qquad \omega_p^2= {n_e(t) e^2\over \veps_i m_*}~,
\eeq
where $\omega_p$ is the plasma frequency and $m_*$ is the effective electron mass and the  conduction electron number density $n_e\propto W_{\rm laser}/{\rm pulse}$.\footnote{In such experiments the idea is to fire repeatedly a GigaHertz or even higher frequency laser at a semi-conducting material, e.g., see \cite{Braggio:2005epl, Naylor:2009qj}. The laser power, $W_{\rm laser}$, per pulse and the number of pulses per wave train determine the number of conduction electrons.}. In this case we find that the dielectric response becomes
\beq
\omega^2 \veps(\omega) = \veps_i(t)\omega^2
-\veps_i(t) \omega_p^2 = \veps_i(t)\omega^2 - {n_e(t) e^2\over m_*}~,
\eeq
where $\veps_i(t)$ denotes the region $i$ in a cavity with time varying part of the dielectric and $n_e(t)$ the time varying part of the conduction electrons. 
The idea of a plasma-mirror (window) was first put forward by \cite{Yablonovitch:1989} to model Hawking and Unruh-like radiation using non-linear properties of a dielectric. However, within the last ten years real experiments to detect DCE radiation have started and are underway, e.g., see \cite{Braggio:2005epl}. As we shall see later on this can mimic the coupling of the inflaton to matter fields, where in Eq. (\ref{Lag}) we assume
\beq
\Psi\equiv \chi, \qquad$and$\qquad
m_p^2(t)= n_e(t)e^2/m_*g^2 \equiv\phi^2(t)= g^2 \Phi^2 \sin^2(mt)
\label{plasmirror}
\eeq
where we can also infer another analogy: $a(t)\equiv\veps_i(t)$, but as we shall see this requires a more complicated experimental set up if we wish to consider (p)reheating at the same time as cosmic expansion. 

\par The separation of variables is not always possible, particularly for time-dependent cases. However, a convenient way to separate Maxwell's equations is to use Hertz vectors; developed by \cite{Nisbet:1957} for non-dispersive inhomogeneous media (also see \cite{Naylor:2013}). In the case of a plasma-mirror the rigorous separation of Maxwell's equations including a time-dependent conductivity will be left as future work. For now we will just assume that the EOM derived from the Lagrangian in Eq. (\ref{toyL}) leads to the following Hertz-like equation for TE-modes:\footnote{As discussed in \cite{Naylor:2012ia}, the conductivity/polarization may only affect TM, not TE modes.}
\beq 
\partial_t ({\veps_i(t)\partial_t  {\mathbf \Pi}_m}) -\nabla^2 { {\mathbf \Pi}_m} + m_p^2(t) \bb\Pi_m= 0~.
\label{Maxwell}
\eeq


\par  From symmetry considerations (e.g., for a rectangular or cylindrical cavity) if $\bb \Pi_{ m}= \Psi~\hat \mathbf z$ then separation of variables (\cite{PedrosaPRL2009}):
\beq
\Psi(\bb x,t)=\sum_{\bb l} \psi_{\bb l}(\bb x) q_{\bb l}(t)~,
\label{sep}
\eeq 
with $\bb l = (n,p,l)$ leads to the following wave equation:
\beq
\bb\nabla^2 \psi_{\bb l}(\bb x)+\veps_i(t)\omega_{i\bb l}^2(t)\psi_{\bb l}(\bb x)=0~.
\label{nabla}
\eeq
 We then find that the time dependent part satisfies
\beq
\ddot {q}_{\bb l} + {\dot\veps_i \over \veps_i} \dot {q}_{\bb l} + \omega_{i\bb l}^2 q_{\bb l}=0
\label{timeQ}
\eeq
where 
\beq
{\omega_{i\bb l}^2(t) } ={c^2\over \veps_i(t)} \left[\left(\frac{x_{\rm np}}{r}\right)^2 +\left( n\pi\over L\right)^2 \right]+ {n_e(t)e^2\over \veps_i(t) m_*}
\label{timeQ}
\eeq 
and we have assumed a TE mode dispersion relation for photons in a cylindrical cavity of radius $r$ where $a\ll L$ in Fig. \ref{exp}, e.g., see \cite{UhlmanPRL2004}. In the above we used the following orthonormality condition:
\beq
\int_{0}^{L} d^3{\bb x}\psi_{\bb l}(\bb x)\psi_{\bb n}(\bb x)=(\psi_{\bb l},\psi_{\bb n})=\delta_{\bb l\bb n}
\label{ortho}
\eeq 
where the bounds with $(0,L)$ are for photons in a cavity of axial length $L$. 

\par To find a separable time-dependent solution, without assuming an instantaneous basis approximation, we rescale the field as $\tilde q_{\bb l} = \veps^{1/2} q_{\bb l}$ to get an equation in Mathieu-like form:
\beq
\ddot {\tilde q}_{ l}  + \tilde\omega_{il}^2(t)\tilde q_{ l}=0
\label{DCEMathieu}
\eeq 
where
\beq
\tilde \omega_{il}^2(t) = \left[\omega_{il}^2+\frac 1 4 {\dot \veps^2\over \veps^2} - \frac 1 2 {\ddot \veps \over \veps} \right] 
\approx {c^2n^2\pi^2\over \veps^2 L^2}+  {n_e(t)e^2\over m_*}
+\frac 1 4 {\dot \veps^2\over \veps^2} - \frac 1 2 {\ddot \veps \over \veps}  
\label{freqDCE}
\eeq
and to keep things simple, we have assumed a photon cavity fundamental mode $x_{01}=2.4048\ll r$ (actually this behaves as an effective mass term if $\veps_i$ is constant). Hence the 3D cavity modes become approximately one-dimensional $\bb l\to l$. 

\par If we now compare this equation with Eq. (\ref{freqP}) for $\xi=1/6$ and $m_\chi=0$, we see that the last two terms in Eq. (\ref{freqDCE}) are equivalent to a massless conformally coupled field (as expected for Maxwell's equations). The plasma mass, $m_p$ breaks the conformal invariance, as would nonlinear perturbations of the dielectric function: $\veps_{NL}=\veps +\delta\veps$, e.g. see \cite{Westerberg:2014}.

\par Interestingly, by comparing the first and second terms of Eq. (\ref{freqP}) with those in Eq. (\ref{freqDCE}), then with units of $c=1$ we see that we can make the analogy:
\beq
\veps_i(t) \equiv a(t),\qquad \Psi\equiv \chi$ \quad and \quad ${m_p^2\over\veps_i}={n_e(t)e^2\over \veps_i m_*}\equiv g^2 \Phi^2\sin^2(m\,t)~. 
\label{analogy}
\eeq
The above analogy is one of the main results of this article: the dielectric behaves like a scale factor for cosmic expansion, while the plasma mass behaves as an oscillating inflaton field (with coupling $g^2$). We will come back to discuss simulation schemes in Sec. \ref{sims}, where for suitable chosen materials and equipment we will argue that (p)reheating can be simulated.\footnote{Actually, in the plasma-mirror model described here, we need two different regions $\veps_I, \veps_{II}$ with $\veps_{II}$ a constant to have a better analogy of (p)reheating with cosmic expansion, see Sec. \ref{sims}.} 


\section{Particle Creation}
\label{create}

\par In the case of time-dependent media, the number of particles created was found via the Bogoliubov method \cite{DodonovPRA47}; however an equivalent method using only mode functions will be used here, e.g., see \cite{Kofman:1997}. 

\par Much like in Sec. \ref{intro}, except now the modes are discrete,\footnote{Most proposed DCE experiments are placed in cavities (and hence the modes are discrete) for the very same reason of parametric enhancement in preheating. In most models of inflation we transit from a potential hill to a trough as we go from inflationary expansion to (p)reheating. The potential well is usually responsible for the parametric enhancement; however, dissipation can also enhance (p)reheating, see \cite{Branden:2014}.} the quantum field operator expansion in the Heisenberg representation for our TE Hertz-like potential, $\Psi$ is
\beq
\hat{\Psi}({\bf x},t) =\sum_{ l} 
\left[ \hat a_{\bb l} \psi_{ l}(\bb x) q_{ l}(t)+ \hat a_{ l}^{\dagger}  \psi^*_{ l}(\bb x) q^*_{ l}(t)
\right]~,
\eeq
where $\hat a_{ l}, \hat a_{ l}^{ \dagger}$ are annihilation and creation operators respectively and the mode functions $ \psi_{\bb l}(\bb x),~ q_{\bb l}(t)$ were defined in Eqs. (\ref{nabla},\ref{ortho},\ref{timeQ}); again we assume $\bb l \to l$. Here we need to impose standard initial conditions relevant to particle creation at $t=0$: 
\beq
q_l(0)={1\over \sqrt{ 2\tilde\omega_l}}$\qquad and \qquad$\dot q_l(0)=-i\sqrt{\tilde\omega_l\over 2}~.
\eeq
The particle number density can be obtained directly from the energy of each mode divided by energy $\omega_{\bb l}$ of each particle (\cite{Kofman:1997}):
\beq
n_{l} = {\omega_l\over 2} \left( 
{|\dot{\tilde q}_l|^2\over \omega_l^2}+ |\tilde q_l|^2
\right) -\frac 1 2
\label{numbden}
\eeq
where in the above expression we have subtracted off the zero point energy (units $\hbar, c=1$). Equations (\ref{PMathieu}, \ref{DCEMathieu}) are in Mathieu form and have a well known structure of narrow or broad band resonances for certain parameters \cite{MathieuBook}. We stress that this method has separated variables without using an instantaneous basis approximation, cf. \cite{Yamamoto:2011pra}.

\par Before numerically solving for the number of created particles, we can estimate the pair creation rate in a method slightly different to the theory of the Mathieu equation, e.g., see \cite{Kofman:1997}. If the background field (the laser) leads to shifts in frequency near to parametric resonance:  
\beq
\omega_l^2 (t)\sim \omega_{0l}^2+\Delta  \omega_{0l}^2= \omega_{0l}^2(1+ \kappa \cos(\Omega_{l} t)),
\label{para}
\eeq
where the driving frequency, $\Omega_l$, is chosen as  $\Omega_{l}=2\omega_{0l}$, then in the late time limit it is possible to show  \cite{DodonovPRA47}:
\beq
n_{ l} \approx \sinh^2 \left({\omega_{0l} \kappa  t /4 } \right) \simeq
\frac 1 4 e^{\frac 1 2 \omega_{0l}\kappa t}~.
\eeq
This implies that for given parameters we have a decay rate $\propto\dot n_{l}/\ n_l$ and so the number of $\chi$ particles (or photons in our analogy) depends on the previous number of $\chi$ particles. Comparing to the notation used in \cite{Kofman:1997}, we have \beq
\omega_{0l}\kappa=q m~,
\eeq
where the value of $q$ determines the type of resonance: narrow ($q\ll 1$) or broad ($q \gg 1$).

\par Comparing this rate with that in Eq. (\ref{pertG}), and writing Eq. (\ref{freqP}) in the form of Eq. (\ref{para}) we see that for preheating:
\beq
\dot n_l/n_l = g^2\Phi^2 /(4m)
\eeq
and this is larger than perturbative reheating, c.f. Eq. (\ref{pertG}),   for $\Phi > g^2 m^2/(8\pi)$ (as was originally discussed in \cite{Kofman:1997}). In the perturbative limit it was proportional to $g^4$, while now it is $g^2$ and hence has a larger rate in the weak coupling regime.


\section{Simulation Scheme}
\label{sims}

\par
\begin{figure}[t]
\centering
\scalebox{0.45}{\includegraphics{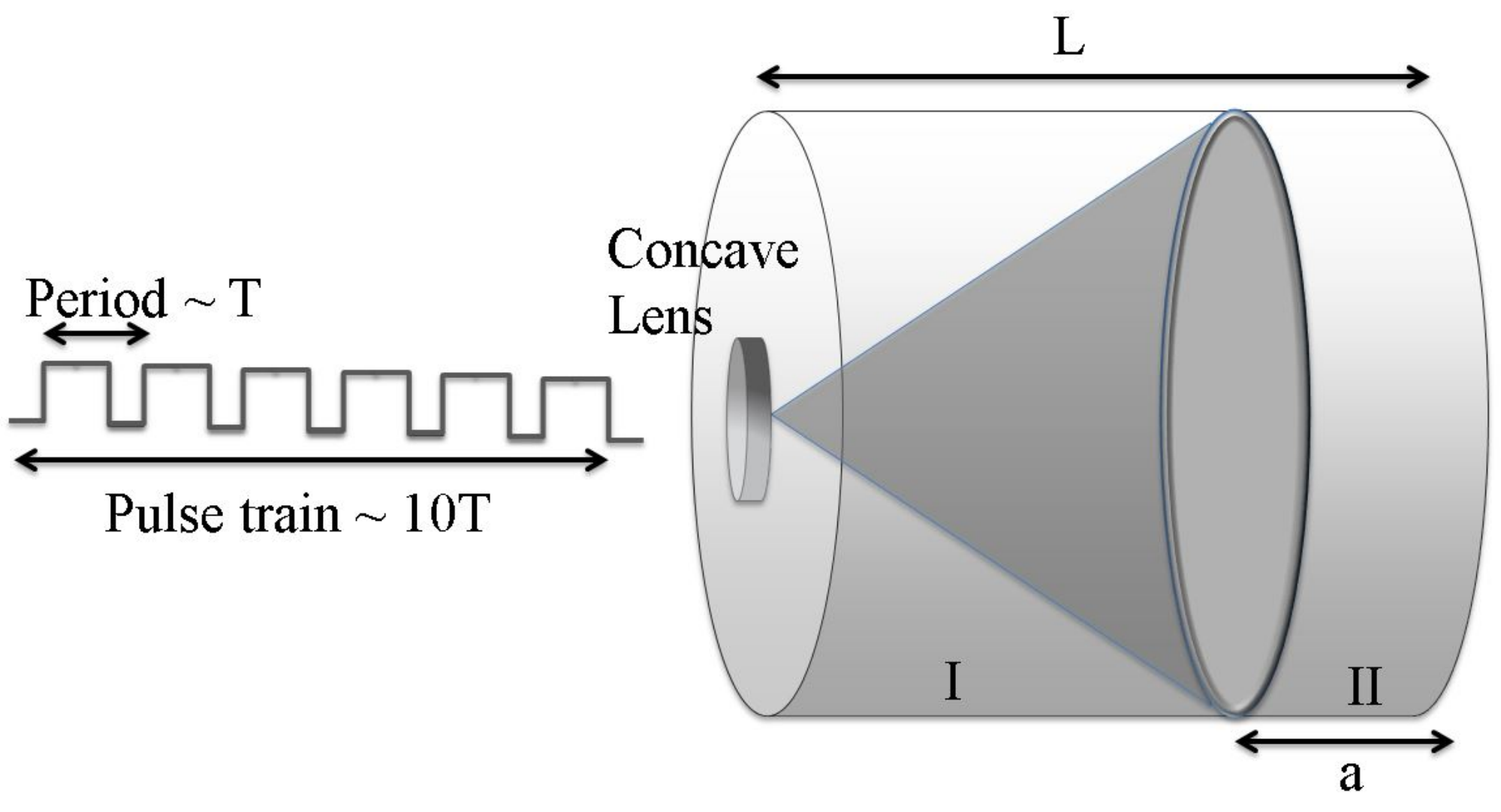}}
\caption{We assume that $n_e(t)e^2/(\veps_{II}m_*)\equiv g^2 \Phi^2 \sin^2(mt) $ for  inflaton field oscillations, where in region $II$ we assume $\veps_{II}=const$ and $n_e(t)\approx n_{max} \sin^2(t)$ in units of $1/(2\pi T)$ for pulse period $T$. For simplicity we show a square-wave for the laser pulse tuned to a wavelength that excites valence electrons to the conduction band, creating a plasma-mirror (typically $a\ll L$ or $L-a\ll L$). 
}
\label{exp}
\end{figure}

  We are now in a position to observer how DCE experiments could be used to simulate (p)reheating in inflation methods via the modulation of time-dependent properties in quantum optics. One such proposal by  \cite{Braggio:2005epl} (also see \cite{Naylor:2009qj}) is based on the plasma-mirror model (\cite{Yablonovitch:1989}) which is an experiment designed to detect dynamical Casimir effect (DCE) radiation by the modulation of the electron conduction density $n_e(t)$. In Fig.~\ref{exp} the general idea is sketched, where a pulsed laser of an appropriate frequency (period $T$) can be used to vary $n_e(t)$. The exact details of the shape of the pulse profile, $n_e(t)$ can be found in \cite{Naylor:2009qj}, but in practice a $sin^2$ (or Gaussian-like) laser profile, will lead to an asymmetric Gaussian profile for the plasma-mirror.

\par In typical proposed plasma-mirror experiments we would place a thin sheet of material to excite {\it surface} plasma electrons with $n_s(t)\approx n_e d_s$, where $d_s$ is the penetration depth, somewhere inside the cavity (\cite{Naylor:2009qj}). Here we just assume that for very thin sheets we can use a delta-function approximation for the mass term: $m_p^2=\delta(z-a) m_s^2$, cf. Eq. (\ref{toyL}). As a further simplification, in Fig. \ref{exp} we have assumed only two regions ($I$ and $II$), but in the limit that $a\ll L$ then we have a very thin ``plasma-sheet" at one end of the cavity, which is the usual choice for current experiments to detect DCE radiation with plasma sheets \cite{Agnesi:2008ja}. Alternatively, and the simple choice we will use in this article, is for $L-a \ll L$ which is for a cavity almost completely filled with dielectric $\veps_{II}$.

\par Hence, given only very few modifications (in \cite{Agnesi:2008ja} a rectangular cavity is used) if a DCE signal can be found in the sense of measuring $n(t)$ at given time steps, then we might be able to simulate (p)reheating after inflation. The only real difference (but not such a hindrance) would be to try and construct a long narrow cylindrical/rectangular cavity to reduce the transverse mode terms: $(x_{np}/r)^2$, which behave as an effective $m_\chi^2$ mass term for $\veps_{II}=const.$ and to have $n\pi/L \to k$ to have continuous frequencies in Eq. (\ref{freqDCE} ) for the eigenfrequencies.

\par It is important to mention that the analogy with (p)reheating during cosmic expansion is only really correct if $m_p^2=n_e e^2/(\veps_{II} m_*)$ for $\veps_{II}=const$. Then, in order to have mode functions that vary with an analogue scale-factor we need to have a different region,  
$\veps_{I}(t)$, to mimic the effect of cosmic expansion. This could be achieved by including a extra laser, with different wavelength (not shown in Fig. \ref{exp}) tuned to excite Rabi oscillations, rather than excite valence electrons to the conduction band.  If the profile of the amplitude increases as some power, e.g.,  $\veps_{I}\propto t^{2/3}$, or modulates on longer time scales with some power of $t$, then we might even be able simulate preheating while including expansion effects. 

\begin{figure}[t]
\centering
\scalebox{0.85}{\includegraphics{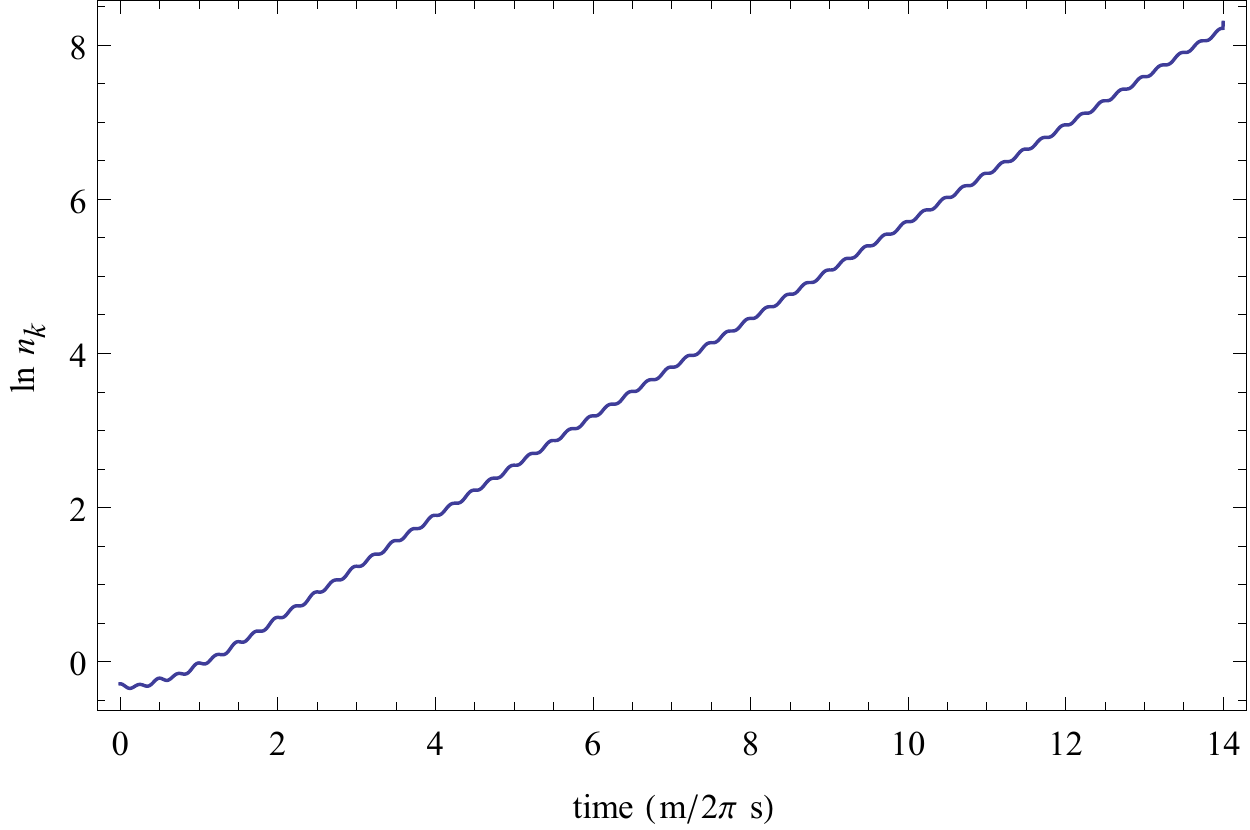}}
\scalebox{0.85}{\includegraphics{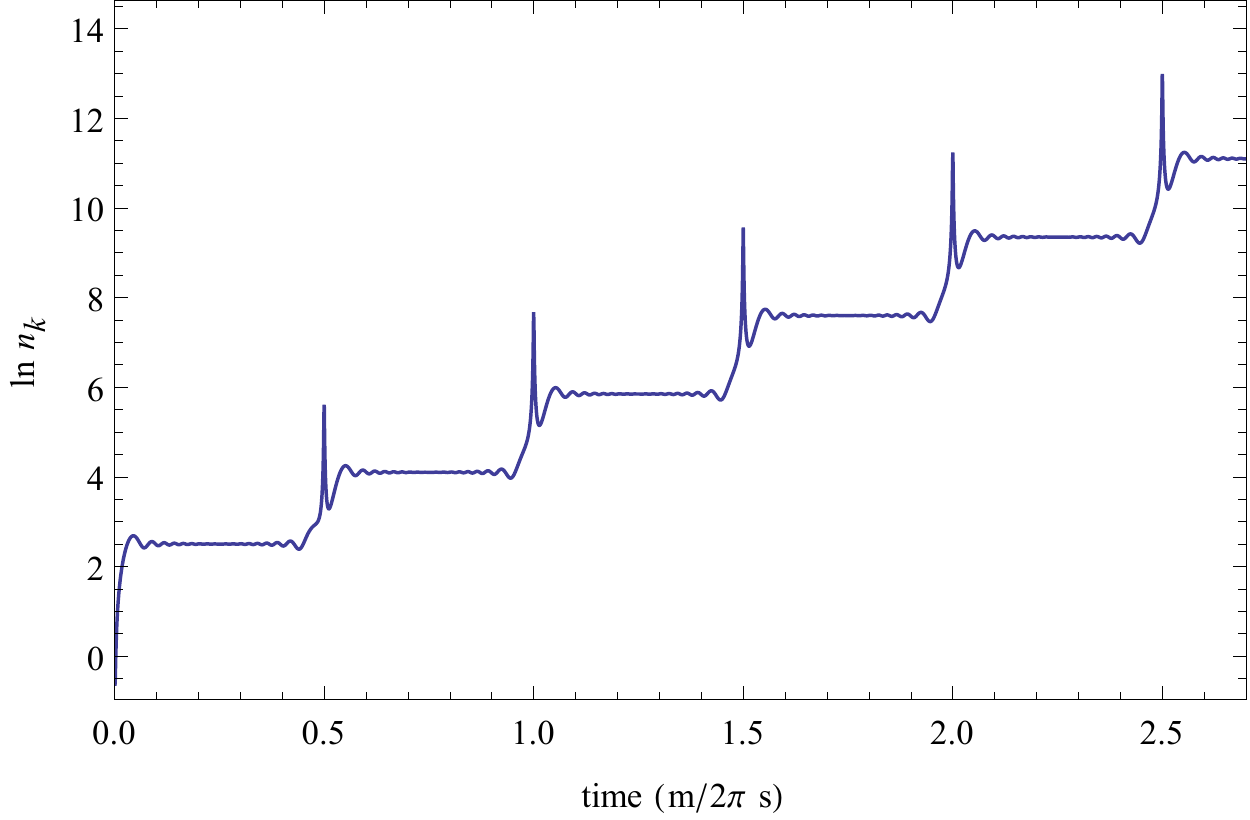}}
\caption{Typical example of narrow \& broad parametric resonance for $\log n_k(t)$ in (p)reheating/DCE, where time is in units of $m/(2\pi)\equiv 1/(2\pi T)$, see Fig. \ref{exp}. For simplicity in this analogue experiment we have $L-a\ll L$. Upper panel: narrow band values are for $m=1.125, k=1$ and $q= g^2\Phi^2/(8m^2)\sim 0.1$. Lower panel: broad band values are for $m=10, k=1$ and $q= g^2\Phi^2/(8m^2)\sim 200$.
}
\label{res}
\end{figure}

\par To complete our analogy we briefly look at the values used in Fig. \ref{res}, for narrow and broad resonance regimes, cf. Fig. 2 and 3 of  \cite{Kofman:1997}. Assuming the bulk pulse profile is of the form $n_e(t) = n_{max} \sin^2(t)$ with period $T$ in a pulse (see Fig. \ref{exp}) in units of $1/(2\pi T)$, then we have a simple analogy with (p)reheating: $1/T \equiv m$ (the inflaton mass) and if $n\pi/(\veps_{II} L)\to k\sim 1$, then $ n_{max}e^2/(\veps_{II}m_*) \equiv qm  = V_{max}$ in units of ${\rm m^{-1}}$. We now assume that the range of $V_{max}$  can be controlled via the laser power and number of pulses per wave train, \cite{Naylor:2009qj}: $V_{max} \sim 0 - 10^5\,{\rm m^{-1}}$, for $\veps_{II} \sim {\cal O}(1)$. Therefore, for large laser powers ($V_{max}\gg 1$) we get broad resonance, while for low laser powers ($V_{max}\ll 1$) we obtain narrow resonance.

\section{Discussion \& Outlook}

\par In summary we have discussed some close similarities between the DCE in quantum optics and preheating in inflationary cosmology; albeit with what might be some rather ad-hoc assumptions: we have assumed the plasma-mirror model conductivity is for the TE modes in Maxwell's equations and not proven that they separate for a time-dependent conductivity. The shape of the laser pulse and the response of the dielectric should also be considered. Our claim about simulating cosmic expansion during (p)reheating was also based on the assumption of two different regions, and we have not solved for the eigenfrequencies in this case. Actually, in Fig. \ref{res}, we have plotted $\log n(t)$ under the assumption of an almost completely filled cavity, $L-a\ll L$. The effect of simulated expansion would modify the eigenvalues in Eq. (\ref{freqDCE}). We hope to address these issues in forthcoming work.

\par An observant reader might ask if the model we have considered here (the plasma-mirror model) really has some analogy to that in cosmological (p)reheating? For example, the motion of a pendulum also undergoes parametric oscillations and hence obeys a Mathieu-like equation (as does any vibrational system). However, both the DCE and (p)reheating arise from non-adiabatic changes in the quantum vacuum state, which is very different to the notion of the classical vibrations of a system, e.g., see \cite{Nation:2012}.

\par Furthermore, doesn't the DCE arise from photon-pair creation, while scalar preheating arises from two particle processes $\phi^2\to \chi^2$ (in our toy Lagrangian, Eq. (\ref{toyL}))? When we consider a moving boundary or plasma-mirror we have assumed a classical background for the electrons involved, whereas in practice these can also be modeled microscopically using for example the Hopfield Lagrangian. This was first used in the context of the DCE by \cite{Shimizu_JJAP1995} (also see references in \cite{Westerberg:2014}). 

\par There are some close parallels with \cite{Schutz:2007} who considered the simulation of particle creation from cosmic expansion in ion traps; however, here we are looking at (p)reheating (with or without cosmic expansion) and using a physically different simulation apparatus. There are also analogs of the DCE in Bose-Einstein condensates in ion-traps, e.g., see \cite{Dodonov:2013} and references therein, which are also possible candidates for simulating (p)reheating.  Interestingly, an alternative analogue DCE effect has already detected evidence for the DCE, via vacuum squeezing, in metamaterial quantum-circuits using waveguides \cite{Paraoanu_PNAS110}. This leads to the tantalizing fact that we may already be able observe the signature of analogue (p)reheating in these set ups. 

\par Finally, to mention just a few of many other interesting avenues of research: we might be to try and simulate the quantum to classical transition during inflation (if at all possible), to verify if preheating leads to reheating as time evolves and to look at (p)reheating when the speed of sound is less than $c$, such as in k-inflation models, e.g. ,see  \cite{Mukhanov:2005}. Hopefully we have argued how knowledge from time-dependent effects in quantum optics, such as in the DCE, can be applied to (p)reheating and vice versa. 

\section*{Acknowledgements}

I would like to thank initial discussions on this topic with Juan Garcia-Bellido (Universidad Aut\'onoma de Madrid) relating to the possibility of simulating preheating in the DCE and in Bose-Einstein condensates. Much of the work here relating to (p)reheating is based on the Osaka University Master's Thesis (2014) of Hiroaki Tagawa (Fujitsu); I am indebted to him for the adaptation of some of his numerical code.

\bibliographystyle{abbrvnat}
\bibliography{ref}


\end{document}